# Power Assignment Problems in Wireless Communication

Stefan Funke  Sören Laue  Rouven Naujoks         Zvi Lotker

Max-Planck-Institut f. Informatik         Ben Gurion University
Saarbrücken, Germany                       Beer Sheva, Israel
{funke,laue,naujoks}@mpi-inf.mpg.de       zvilo@cse.bgu.ac.il

**Abstract**

A fundamental class of problems in wireless communication is concerned with the assignment of suitable transmission powers to wireless devices/stations such that the resulting communication graph satisfies certain desired properties and the overall energy consumed is minimized. Many concrete communication tasks in a wireless network like broadcast, multicast, point-to-point routing, creation of a communication backbone, etc. can be regarded as such a power assignment problem.

This paper considers several problems of that kind; for example one problem studied before in [1, 6] aims to select and assign powers to $k$ of the stations such that all other stations are within reach of at least one of the selected stations. We improve the running time for obtaining a $(1+\epsilon)$-approximate solution for this problem from $n^{((\alpha/\epsilon)^{O(d)})}$ as reported by Bilo et al. ([6]) to $O\left(n + \left(\frac{k^{2d+1}}{\epsilon^d}\right)^{\min\{2k,\ (\alpha/\epsilon)^{O(d)}\}}\right)$ that is, we obtain a running time that is *linear* in the network size. Further results include a constant approximation algorithm for the TSP problem under squared (non-metric!) edge costs, which can be employed to implement a novel data aggregation protocol, as well as efficient schemes to perform $k$-hop multicasts.

# 1 Introduction

Wireless network technology has gained tremendous importance in recent years. It not only opens new application areas with the availability of high-bandwidth connections for mobile devices, but also more and more replaces so far 'wired' network installations. While the spatial aspect was already of interest in the wired network world due to cable costs etc., it has far more influence on the design and operation of wireless networks. The power required to transmit information via radio waves is heavily correlated with the Euclidean distance of sender and receivers. Hence problems in this area are prime candidates for the use of techniques from computational geometry.

Wireless devices often have limited power supply, hence the energy consumption of communication is an important optimization criterion. In this paper we use the following simple geometric graph model: Given a set $P$ of $n$ points in $\mathbb{R}^2$, we consider the complete graph $(P, P \times P)$ with edge weight $\omega(p, q) = |pq|^\alpha$ for some constant $\alpha > 1$ where $|pq|$ denotes the Euclidean distance between $p$ and $q$. For $\alpha = 2$ the edge weights reflect the exact energy requirement for free space communication. For larger values of $\alpha$ (typically between 2 and 4), we get a popular heuristic model for absorption effects.

A fundamental class of problems in wireless communication is concerned with the assignment of suitable transmission powers to wireless devices/stations such that (1) the resulting communication graph satisfies a certain connectivity property $\Pi$, and (2) the overall energy assigned to all the network nodes is minimized. Many properties $\Pi$ can be considered and have been treated in the literature before, see [7] for an overview. In this paper we consider several definitions of $\Pi$ to solve the following problems:

**$k$-Station Network/$k$-disk Coverage:** *Given a set $S$ of stations and some constant $k$, we want to assign transmission powers to at most $k$ stations (senders) such that every station in $S$ can receive a signal from at least one sender.*

**$k$-hop Multicast:** *Given a set $S$ of stations, a specific source station $s$, a set of clients/receivers $C \subseteq S$, and some constant $k$, we want the communication graph to contain a directed tree rooted at $s$ spanning all nodes in $C$ with depth at most $k$.*

**TSP under squared Euklidean distance:** *Given a set $S$ of $n$ stations, determine a permutation $p_0, p_1, \ldots p_{n-1}$ of the nodes such that the total energy cost of the TSP tour, i.e. $\sum_{i=0}^{n-1} |p_i p_{(i+1) \bmod n}|^\alpha$ is minimized.*

## 1.1 Related Work

The *k-Station Network Coverage* problem was considered by Bilo et al. [6] as a $k$-disk cover, i.e. covering a set of $n$ points in the plane using at most $k$ disks such that the sum of the areas of the disks is minimized. They show that obtaining an exact solution is $\mathcal{NP}$-hard and provide a $(1+\epsilon)$ approximation to this problem in time $n^{((\alpha/\epsilon)^{O(d)})}$ based on a plane subdivision and dynamic programming. Variants of the $k$-disk cover problem were also discussed in [1].

The general broadcast problem – assigning powers to stations such that the resulting communication graph contains a directed spanning tree and the total amount of energy used is minimized– has a long history. The problem is known to be $\mathcal{NP}$-hard ([8, 7]), and for arbitrary, non-metric distance functions the problem can also not be approximated better than a log-factor unless $\mathcal{P} = \mathcal{NP}$ [13]. For the Euclidean setting in the plane, it is known ([2]) that the minimum spanning tree induces a power assignment for broadcast which is at most 6 times as costly as the optimum solution. This bound for a MST-based solution is tight ([8], [14]). There has also been work on restricted broadcast operations more in the spirit of the *k-hop multicast* problem we consider in this paper. In [3] the authors examine a *bounded-hop* broadcast operation where the resulting communication graph has to contain a spanning tree rooted at the source node $s$ of depth at most $k$. They show how to compute an optimal $k$-hop broadcast range assignment for $k = 2$ in time $O(n^7)$. For $k > 2$ they show how to obtain a $(1 + \epsilon)$-approximation in time $O(n^{O(\mu)})$ where $\mu = (k^2/\epsilon)^{2^k}$, that is, their running time is triply exponential in the number of hops $k$ and this shows up in the exponent of $n$. In very recent work [12], Funke and Laue show how to obtain a $(1 + \epsilon)$ approximation for the $k$-hop broadcast problem in time doubly exponential in $k$ based on a coreset which has size exponential in $k$, though.

The classical travelling salesperson problem is NPO-complete for arbitrary, non-metric distance functions (see [11]), a lot of progress has been made for the geometric case, where a $(1+\epsilon)$ approximation is available (see [4]).

General surveys of algorithmic range assignment problems can be found in [7, 15, 10].

## 1.2 Our Contribution

In Section 2 we show how to find a coreset of size **independent of** $n$ and polynomial in $k$ and $1/\epsilon$ for the $k-$Station Network Coverage/$k$-Disk cover problem. This enables us improve the running time of the $(1 + \epsilon)$ approximation algorithm by Bilo et al.[6] from $n^{((\alpha/\epsilon)^{O(d)})}$ to $O\left(n + \left(\frac{k^{2d+1}}{\epsilon^d}\right)^{\min\{2k,\ (\alpha/\epsilon)^{O(d)}\}}\right)$, that is, we obtain a running time that is *linear* in $n$. We also present a variant that allows for the senders to be placed arbitrarily (not only within the given set of points) as well as a simple algorithm which is able to tolerate few outliers and runs in polynomial time for constant values of $k$ and the number of outliers.

Also based on the construction of a (different) coreset of small size, we show in Section 3 how to obtain a $(1+\epsilon)$ approximate solution to the $k$-hop multicast problem with respect to a constant-size set $C$ of receivers/clients. Different from the solution for the $k$-hop broadcast problem presented in [12] we can exhibit a coreset of size *polynomial* in $k, 1/\epsilon$ and $r$. The approach in [12] requires a coreset of size exponential in $k$.

Finally, in Section 4 we consider the problem of finding energy-optimal TSP tours. The challenge here is that the edge weights induced by the Energy costs do not define a metric anymore; a simple example shows that an optimal solution to the Euklidean TSP can be a factor $\Omega(n)$ off the optimum solution. We present an $O(1)$-approximation for the TSP problem with powers of the Euklidean distance as edge weights.

## 2 Energy-minimal Network Coverage or: "How to cover Points by Disks"

Given a set $S$ of points in $\mathbb{R}^d$ and some constant $k$. We want to find at most $k$ $d$-dimensional balls with radii $r_i$ that cover all points in $S$ while minimizing the objective function $\sum_{i=1}^{k} r_i^\alpha$ for some power gradient $\alpha > 1$.

### 2.1 A small coreset for $k$-disk cover

In this section we describe how to find a coreset of size $O(k^{2d+1}/\epsilon^d)$, i.e. of size independent of $n$ and polynomial in $k$ and in $1/\epsilon$. We can distinguish between two variants of the problem: the discrete version in which all the center points of the balls must be contained in the input set $S$ and the non-discrete version in which the center points can be chosen arbitrarily. In the construction of the coreset we will focus on the latter version and mention when things have to be changed to make the approach also work for the first case.

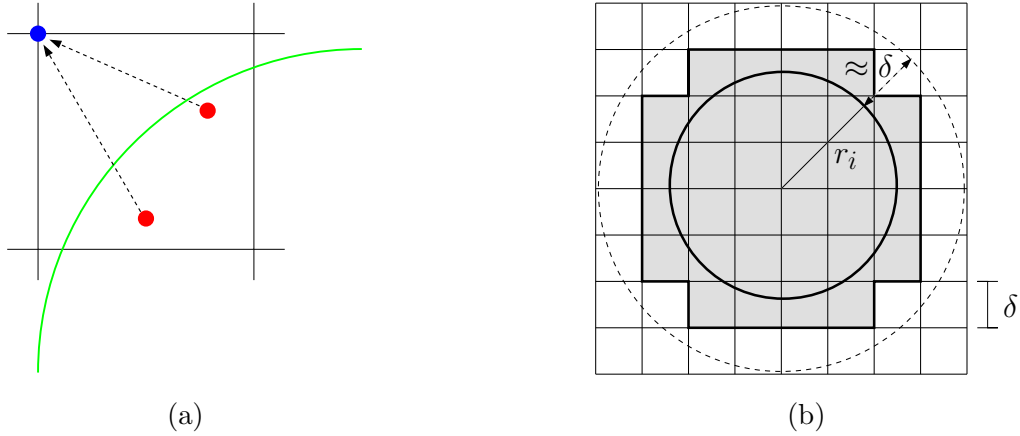

Figure 1: proof illustrations for theorem 2

The idea is to reduce the input size by snapping the input points onto a regular grid in such a way that a feasible solution for the original point set can be transformed into a feasible solution for the aligned point set and vice versa without changing the objective value too much.

We start by putting a regular $d$-dimensional grid on the input set $S$ with grid cell width $\delta$. Each point of $S$ is associated with an arbitrary but fixed corner of the grid cell $C$ in which it is contained. Such a corner point we call the *representative* point for the points in $S \cap C$. We say that $C$ is *active* if $S \cap C \neq \emptyset$. Note that the distance between any point in $S$ and its representative is at most $\sqrt{d} \cdot \delta$ and that from sets of points with same coordinates only one has to be considered. The goal is now to (i) map $S$ to only a few representatives and (ii) the set of representatives $R$ represents $S$ in an $(1+\epsilon)$-approximation manner, i.e. $opt(R) \leq (1+\epsilon) \cdot opt(S)$.

**Theorem 1** *For an appropriate $\delta$ we have*

$$opt(R) \leq (1+\epsilon) \cdot opt(S)$$

**Proof:** Suppose we are given an optimal solution of $S$ by $k$ balls $C_i$ with radii $r_i$ and objective value $OPT$. Now perturb the input points in $S$ by snapping them to the d-dimensional grid as described above. By this perturbation it can happen that some representative points are not covered anymore by the discs $D_i$. Note that increasing the radii $r_i$ by $\sqrt{d} \cdot \delta$ ensures coverage again. ( Notice that in the discrete case the centers are also perturbed. Increasing the radii by just another $\sqrt{d} \cdot \delta$ ensures coverage in this case. ) Thus the cost of this feasible solution $P$ is given by

$$cost(P) = \sum_{i=1}^{k} (r_i + \sqrt{d} \cdot \delta)^\alpha$$

consider one ball $C'_i$ of $P$ and choose

$$\delta := \frac{1}{\sqrt{d}} \cdot \delta' \quad \text{with} \quad \delta' := \frac{\epsilon \cdot OPT^{1/\alpha}}{k \cdot c_\alpha}$$

then we have

$$\begin{aligned} cost(C'_i) &= (r_i + \delta')^\alpha \\ &= \sum_{j=0}^{\alpha} \binom{\alpha}{j} \cdot r_i^\alpha \cdot \delta'^{\alpha-j} = r_i^\alpha + \sum_{j=0}^{\alpha-1} \binom{\alpha}{j} \cdot r_i^\alpha \cdot \delta'^{\alpha-j} \end{aligned}$$

note that $r_i \leq OPT^{1/\alpha}$. Thus

$$cost(C_i') \leq r_i^\alpha + OPT \cdot \underbrace{\sum_{j=0}^{\alpha-1} \binom{\alpha}{j} \cdot \left(\frac{\epsilon}{k \cdot c_\alpha}\right)^{\alpha-j}}_{(*)}$$

using some calculus we can show that $(*) \leq \epsilon/k$ for

$$c_\alpha := \frac{\epsilon}{k} \cdot \frac{1}{(\frac{\epsilon}{k}+1)^{\frac{1}{\alpha}} - 1}$$

If we assume that $\epsilon \leq 1$ (which is reasonable for an approximation scheme) $c_\alpha \leq \frac{1}{\sqrt[\alpha]{2}-1} \leq \ln 2 \cdot \alpha$, i.e. $c_\alpha$ is actually a very small constant, depending only on $\alpha$. Indeed we can show that bounding $c_\alpha$ by $\ln 2 \cdot \alpha$ leads to the asymptotically best bound on the size of the resulting coreset. Thus the cost of $P$ can be bounded by

$$\begin{aligned} cost(P) &\leq \sum_{i=1}^{k} \left(r_i^\alpha + \frac{\epsilon}{k} \cdot OPT\right) \\ &= \sum_{i=1}^{k} r_i^\alpha + \epsilon \cdot OPT = (1+\epsilon) \cdot OPT \end{aligned}$$

∎

**Theorem 2** *The size of the computed coreset $R$ is bounded by*

$$O\left(\frac{k^{2d+1}}{\epsilon^d}\right)$$

**Proof:** Observe that the size of $R$ is exactly given by the number of active cells. On the other hand a cell $C$ is active if and only if there is a point in $S$ that is contained in $C$. Since a feasible solution covers all points, a cell is active only if it is (partially) covered by any of the corresponding balls (see figure 1a). Thus the number of active cells is bounded by the number $\#cc$ of cells (partially) covered by an optimal solution. Thus we can bound $\#cc$ by a simple volume argument: just count how many cells with volume $\delta^d$ fit in the balls of an opimal solution. Note that to ensure that also the partially covered cells are taken into account we increase the radii by $\sqrt{d} \cdot \delta$ (see figure 1b). This leads to

$$\begin{aligned} \#cc &\leq \sum_{i=1}^{k} \frac{\left(r_i + \sqrt{d} \cdot \delta\right)^d}{\delta^d} \\ &\leq \frac{1}{\delta^d} \cdot \sum_{i=1}^{k} \left(OPT^{1/\alpha} + \delta'\right)^d = \frac{1}{\delta^d} \cdot \sum_{i=1}^{k} (OPT^{1/\alpha} + \underbrace{\frac{\epsilon}{k \cdot c_\alpha}}_{\leq 1} OPT^{1/\alpha})^d \\ &\leq \frac{1}{\delta^d} \cdot \sum_{i=1}^{k} (2 \cdot OPT^{1/\alpha})^d \\ &\leq \frac{k^{d+1}}{\epsilon^d} \cdot (2 \cdot c_\alpha \cdot \sqrt{d})^d \quad \in \quad O\left(\frac{k^{d+1}}{\epsilon^d}\right) \end{aligned}$$

For the construction of the grid we have assumed so far that we know the optimal objective value $OPT$. Since we cannot compute $OPT$ efficiently we have to use an approximate value instead. Badoiu et al. show in [5] how to compute a constant factor approximation of the so called $k$-center clustering problem in $O(n)$ time. This problem differs from the $k$-disc cover problem only in the objective function where one pays only for the heaviest disc. Obviously such a solution is a $k$-factor approximation for the $k$-disc cover problem. It is easy to verify that our algorithm is still correct using this approximation instead of $OPT$ but also involves an increase in the size of the computed coreset to $O\left(\frac{k^{2d+1}}{\epsilon^d}\right)$. As future work one could think of deriving a constant factor approximation algorithm to avoid this. ∎

## 2.2 Algorithms

What remains is to solve the small coreset instances. As mentioned before we distinguish between two variants of the problem: the discrete version and the non-discrete version:

### 2.2.1 Discrete Version

**Via Bilo et al.:** Recall that the running time of the approach of Bilo et al. is given by

$$n^{((\alpha/\epsilon)^{O(d)})}$$

using their algorithm for solving the coreset instance yields an overall running time of

$$O\left(n + \left(\frac{k^{2d+1}}{\epsilon^d}\right)^{(\alpha/\epsilon)^{O(d)}}\right).$$

**Via Brute-Force:** We can find an optimal solution in the following way. We consider all $k$-subsets of the points in the coreset $N$ as the possible centers of the balls. Note that at least one point in $N$ has to lie on the boundary of each ball in an optimal solution (otherwise you could create a better solution by shrinking a ball). Thus the number of possible radii for each ball is bounded by $n - k$. In total there are $(n-k)^k \cdot \binom{n}{k} \leq n^{2k}$ possible solutions which means that we can solve our coreset via brute-force in time

$$O\left(n + \left(\frac{k^{2d+1}}{\epsilon^d}\right)^{2k}\right)$$

So we obtain the following result:

**Corollary 1** *The runnnig time of our approximation algorithm in the discrete case is*

$$O\left(n + \left(\frac{k^{2d+1}}{\epsilon^d}\right)^{\min\{\,2k,\ (\alpha/\epsilon)^{O(d)}\,\}}\right)$$

### 2.2.2 Non-Discrete Version

**Via Brute-Force:** Note that on each ball $D$ of an optimal solution there must be at least three points (or two points in diametral position) that define $D$ - otherwise it would be possible to obtain a smaller solution by shrinking $D$. Thus for obtaining an optimal solution via brute force it is only necessary to check all $k$-sets of 3- respectively 2-subsets of $S$ which yields a running time of $O(n^{3k})$. Solving our coreset via brute-force yields the following:

**Corollary 2** *The runnnig time of our approximation algorithm in the non-discrete case is*

$$O\left(n + \left(\frac{k^{2d+1}}{\epsilon^d}\right)^{3k}\right)$$

## 2.3 $k$-disk cover with few outliers

Assume we want to cover not all points by disks but we relax this constraint and allow a few points not to be covered, i.e. we allow let's say $c$ outliers. This way, the optimal cover might have a considerably lower power consumption/cost.

Conceptually, we think of a $k$-disk cover with $c$ outliers as a $(k+c)$-disk cover with $c$ disks having radius 0. Doing so, we can use the same coreset construction as above, replacing $k$ by $k+c$. Obviously, the cost of an optimal solution to the $(k+c)$-disk cover problem is a lower bound for the $k$-disk cover with $c$ outliers. Hence, the imposed grid might be finer than actually needed. So snapping each point to its closest representative still ensures a $(1+\epsilon)$-approximation. Constructed as above, the coreset has size $O(\frac{(k+c)^{2d+1}}{\epsilon^d})$.

Again, there are two ways to solve this reduced instance, first by a slightly modified version of the algorithm proposed by Bilo et al. [6] and second by exhaustive search.

We will shortly sketch the algorithm by Bilo et al. [6] which is based on a hierarchical subdivision scheme proposed by Erlebach et al. in [9]. Each subdivision is assigned a level and they together form a hierarchy. All possible balls are also assigned levels depending on their size. Each ball of a specific level has about the size of an $\epsilon$-fraction of the size of the cells of the subdivision of same level. Now, a cell in the subdivision of a fixed level is called relevant if at least one input point is covered by one ball of the same level. If a relevant cell $S'$ is included in a relevant cell $S$ and no larger cell $S''$ exists that would satisfy $S' \subseteq S'' \subseteq S$, then $S'$ is called a child cell of $S$ and $S$ is called the parent of $S'$. This naturally defines a tree. It can be shown that a relevant cell has at most a constant number of child cells (the constant only depending on $\epsilon$, $\alpha$ and $d$). The key ingredient for the algorithm to run in polynomial time is the fact that there exists a nearly optimal solution where a relevant cell can be covered by only a constant number of balls of larger radius. The algorithm then processes all relevant cells of the hierarchical subdivision in a bottom-up way using dynamic programming. A table is constructed that for a given cell $S$, a given configuration $P$ of balls having higher level than $S$ (i.e. large balls) and an integer $i \leq k$ stores the balls of level at most the level of $S$ (i.e. small balls) such that all input points in $S$ are covered and the total number of balls is at most $i$. This is done for a cell $S$ by looking up the entries of the child cells and iterating over all possible ways to distribute the $i$ balls among them.

The $k$-disk cover problem with $c$ outliers exhibits the same structural properties as the $k$-disk cover problem without outliers. Especially, the local optimality of the global optimal solution is preserved. Hence, we can adapt the dynamic programming approach of the original algorithm. In order for the algorithm to cope with $c$ outliers we store not only one table for each cell but $c+1$ such tables. Each such table corresponds to the table for a cell $S$ where $0, 1, \ldots, c$ pointsx are not covered. Now, we do not only iterate over all possible ways to distribute the $i$ balls among its child cells but also all ways to distribute $l \leq c$ outliers. This increases the running time to $n^{((\alpha/\epsilon)^{O(d)})} \cdot c^{((\alpha/\epsilon)^{O(d)})} = n^{((\alpha/\epsilon)^{O(d)})}$. Hence running the algorithm on the coreset yields the following result:

**Corollary 3** *We can compute a minimum $k$-disk cover with $c$ outliers $(1+\epsilon)$ approximately in*

*time*

$$O\left(n + \left(\frac{(k+c)^{2d+1}}{\epsilon^d}\right)^{(\alpha/\epsilon)^{O(d)}}\right).$$

For the exhaustive search approach we consider all assignments of $k$ disks each having a representative as its center and one lying on its boundary. For each such assignment we check in time $O(kn)$ whether the number of uncovered points is at most $c$. We output the solution with minimal cost.

**Corollary 4** *We can compute a minimum $k$-disk cover with $c$ outliers $(1 + \epsilon)$ approximately in time*

$$O\left(n + k\left(\frac{(k+c)^{2d+1}}{\epsilon^d}\right)^{2k+1}\right).$$

## 3 Bounded-hop Multicast or: "Reaching few Receivers quickly"

Given a set $P$ of points (stations) in $\mathbb{R}^d$, a distinguished source point $s \in P$ (sender), and a set $C \subset P$ of client points (receivers) we want to assign distances/ranges $r : P \to \mathbb{R}_0^+$ to the elements in $P$ such that the resulting communication graph contains a tree rooted at $s$ spanning all elements in $C$ and with depth at most $k$ (an edge $(p, q)$ is present in the communication graph iff $r(p) \geq |pq|$). Goal is to minimize the total assigned energy $\sum_{p \in P} r(p)^\alpha$. This can be thought of as the problem of determining an energy efficient way to quickly (i.e. within few transmissions) disseminate a message or a datastream to a set of few receivers in a wireless network.

As in the previous Section we will solve this problem by first deriving a coreset $S$ of size independent of $|P| = n$ and then invoking a brute-force algorithm. We assume both $k$ and $|C| = c$ to be (small) constants. The resulting coreset will have size *polynomial* in $1/\epsilon$, $c$ and $k$. For few receivers this is a considerable improvement over the exponential-sized coreset that was used in [12] for the $k$-hop broadcast.

### 3.1 A small coreset for $k$-hop multicast

In the following we will restrict to the planar case in $\mathbb{R}^2$, the approach extends in the obvious way to higher (but fixed) dimensions. Assume w.l.o.g. that the maximum distance of a point $p \in P$ from $s$ is exactly 1. We place a square grid of cell width $\Delta = \frac{1}{\sqrt{2}}\frac{\epsilon}{kc}$ on $[-1, 1] \times [-1, 1] \subset \mathbb{R}^2$. The size of this grid is $O(\frac{(kc)^2}{\epsilon^2})$. Now we assign each point in $P$ to its closest grid point. Let $S$ be the set of grid points that had at least one point from $P$ snapped to it, $C'$ the set of grid points that have at least one point from $C$ snapped to it.

It remains to show that $S$ is indeed a coreset. We can transform any given valid range assignment $r$ for $P$ (wrt receiver set $C$) into a valid range assignment $r'$ for $S$ (wrt receiver set $C'$). We define the range assignment $r'$ for $S$ as

$$r'(p') = \max_{p \text{ was snapped to } p'} r'(p) + \sqrt{2}\Delta.$$

Since each point $p$ is at most $\frac{1}{\sqrt{2}}\Delta$ away from its closest grid point $p'$ we certainly have a valid range assignment for $S$. It is easy to see that the cost of $r'$ for $S$ is not much larger than the cost

of $r$ for $P$. We have:

$$\sum_{p' \in S} (r'(p'))^\alpha = \sum_{p \in P} (\max_{p \text{ was snapped to } p'} r(p) + \sqrt{2}\Delta)^\alpha$$
$$\leq \sum_{p \in P} (\max_{p \text{ was snapped to } p'} r(p) + \frac{\epsilon}{kc})^\alpha$$
$$\leq \sum_{p \in P} (r(p) + \frac{\epsilon}{kc})^\alpha.$$

The relative error satisfies

$$\frac{\text{cost}(r')}{\text{cost}(r)} \leq \frac{\sum_{p \in P} (r(p) + \frac{\epsilon}{kc})^\alpha}{\sum_{p \in P} (r(p))^\alpha}.$$

Notice, that $\sum_{p \in P} r(p) \geq 1$ and $r$ is positive for at most $kc$ points $p$ (each of the $c$ receivers must be reached within $k$ hops). Hence, the above expression is maximized when $r(p) = \frac{1}{kc}$ for all points $p$ that are assigned a positive value. Thus

$$\frac{\text{cost}(r')}{\text{cost}(r)} \leq \frac{(kc) \cdot (\frac{1}{(kc)} + \frac{\epsilon}{(kc)})^\alpha}{(kc) \cdot (\frac{1}{(kc)})^\alpha} = (1 + \epsilon)^\alpha.$$

On the other hand we can transform any given valid range assignment $r'$ for $S$ into a valid range assignment $r$ for $P$ as follows. We select for each grid point $g \in S$ one representative $g_P$ from P that was snapped to it. For the grid point to which $s$ (the source) was snapped we select $s$ as the representative. If we define the range assignment $r$ for $P$ as $r(g_P) = r'(g) + \sqrt{2}\Delta$ and $r(p) = 0$ if $p$ does not belong to the chosen representatives, then $r$ is a valid range assignment for $P$ because every point is moved by at most $\Delta/\sqrt{2}$. Using the same reasoning as above we can show that $\text{cost}(r) \leq (1+\epsilon)^\alpha \text{cost}(r')$. In summary we obtain the following theorem:

**Theorem 3** *For the $k$-hop multicast problem with $c$ receivers there exists a coreset of size $O(\frac{(kc)^2}{\epsilon^2})$.*

## 3.2 Solution via a naive algorithm

As we are not aware of any algorithm to solve the $k$-hop multicast problem we employ a naive brute-force strategy, which we can afford since after the coreset computation we are left with a 'constant' problem size. Essentially we consider every $kc$-subset of $S$ as potential set of senders and try out the $|S|$ potential ranges for each of the senders. Hence, naively there are at most $\binom{\frac{kc^2}{\epsilon^2}}{kc} \cdot \left(\frac{kc^2}{\epsilon^2}\right)^{kc}$ different range assignments to consider at all. We enumerate all these assignments and for each of them check whether the range assignment is valid wrt $c'$; this can be done in time $|S|$. Of all the valid range assignments we return the one of minimal cost.

Assuming the floor function a coreset $S$ for an instance of the $k$-hop multicast problem can be constructed in linear time. Hence we obtain the following corollary:

**Corollary 5** *A $(1 + \epsilon)$-approximate solution to the $k$-hop multicast problem on $n$ points in the plane can be computed in time $O(n + \left(\frac{kc}{\epsilon}\right)^{4kc})$.*

As we are only after an approximate solution, we do not have to consider all $|S|$ potential ranges but can restrict to essentially $O(\log_{1+\epsilon} \frac{kc}{\epsilon})$ many, the running time of the algorithm improves accordingly:

**Corollary 6** *A $(1+\epsilon)$-approximate solution to the k-hop multicast problem on n points in the plane can be computed in time $O\left(n + \left(\frac{(kc)^2 \log \frac{kc}{\epsilon}}{\epsilon^3}\right)^{kc}\right)$.*

## 4 Information aggregation via energy-minimal TSP Tours

While early wireless sensor networks (WSNs) were primarily data collection systems where sensor readings within the network are all transferred to a central computing device for evaluation, current WSNs perform a lot of the data processing *in-network*. For this purpose some nodes in the network might be interested in periodically *collecting* information from certain other nodes, some nodes might want to *disseminate* information to certain groups of other nodes. A typical approach for data collection and dissemination as well as for data aggregation purposes are tree-like subnetwork topologies, they incur certain disadvantages with respect to load-imbalance as well as non-obliviousness to varying initiators of the data collection or dissemination operation, though. Another, very simple approach could be to have a *virtual token* floating through the network (or part thereof). Sensor nodes can attach data to the token or read data from the token and then hand it over to the next node. Preferably the token should not visit a node again before all other nodes have been visited and this should happen in an energy-optimal fashion, i.e. the sum of the energies to hand over the token to the respective next node should be minimized. Such a scheme has some advantages: first of all none of sensor nodes plays a distinguished role – something that is desirable for a system of homogenous sensor nodes – furthermore every sensor node can use the same token to initiate his data collection/dissemination operation. Abstractly speaking we are interested in finding a *Travelling Salesperson tour* (TSP) of minimum energy cost for (part of) the network nodes. Unfortunately, the classical TSP with non-metric distance function is very hard to solve (see [11]), most progress has been made for the metric and geometric case (see e.g. [4]).

In this Section we show that the 'normal' Euklidean TSP is not suitable for obtaining an energy-efficient tour, but still a constant-factor approximation can be obtained.

### 4.1 Why Euclidean TSP does not work

Simply computing an optimal tour for the underlying Euclidean instance does not work. The cost for such a tour can be a factor $\Omega(n)$ off from the optimal solution for the energy-minimal tour. Consider the example where $n$ points lie on a slightly bent line and each point having distance 1 to its right and left neighbor. An optimal Euclidean tour would visit the points in their linear order and the go back to the first point. Omitting the fact that the line is slightly bent this tour would have a cost of $(n-1) \cdot 1^2 + (n-1)^2 = n(n-1)$ if the edge weights are squared Euclidean distances. However, an optimal energy-minimal tour would have a cost of $(n-2) \cdot 2^2 + 2 \cdot 1^2 = 4(n-1)+2$. This tour would first visit every second point on the line and on the way back all remaining points as in figure 2.

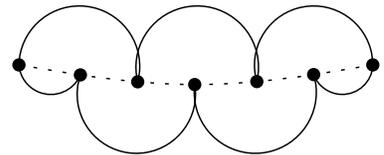

Figure 2: An optimal energy-minimal tour for points on a line

### 4.2 A 6-Approximation Algorithm

In this section we will describe an algorithm which computes a 6-approximation for the TSP under squared Euclidean distance. Obviously, the cost of a minimum spanning tree is a lower bound for the optimal value OPT of the tour.

Consider a non-trivial minimum spanning tree $T$ for a graph with node set $V$ and squared Euclidean edge weights. We denote the cost of such a tree by $\mathrm{MST}(T)$. Let $r$ be the root of $T$ and $p$ be one child of $T$.

We define two Hamiltonian paths $\pi^a(T)$ and $\pi^b(T)$ as follows. Let $\pi^a(T)$ be a path starting at $r$, finishing at $p$ that visits all nodes of $T$ and the cost of this path is at most $6\,\mathrm{MST}(T) - 3\|rp\|^2$. Let $\pi^b(T)$ be defined in the same way but in opposite direction, i.e. it starts at $p$ and finishes at $r$.

Now, if we have such a tour $\pi^a(T)$ for the original vertex set $V$ we can construct a Hamilton tour by connecting $r$ with $p$. The cost of this tour is clearly at most $6\,\mathrm{MST}(T) - 3\|rp\|^2 + \|rp\|^2 \leq 6\,\mathrm{MST}(T) \leq 6\,\mathrm{OPT}$. It remains to show how to construct such tours $\pi^a$ and $\pi^b$. We will do this recursively.

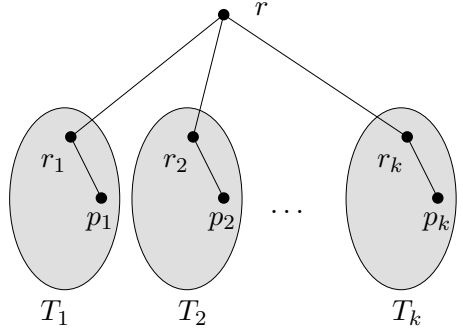

Figure 3: tree $T$ and its children trees $T_1, T_2, \ldots, T_k$

For a tree $T$ of height 1, i.e. a single node $r$, $\pi^a(T)$ and $\pi^b(T)$ both consist of just the single node. Conceptually, we identify $p$ with $r$ in this case. Obviously, the cost of both paths is trivially at most $6\,\mathrm{MST}(T) - 3\|rp\|^2$.

Now, let $T$ be of height larger than 1 and let $T_1, \ldots, T_k$ be its children trees. Let $r$ denote the root of $T$ and $r_i$ the root of $T_i$ and $p_i$ be a child of $T_i$ as in figure 3. Then we set $\pi^a(T) = (r, \pi^b(T_1), \pi^b(T_2), \ldots, \pi^b(T))$.

The cost of the path $\pi^a(T)$ satisfies

$$
\begin{aligned}
\mathrm{cost}(\pi^a(T)) &= \|rp_1\|^2 + \mathrm{cost}(\pi^b(T_1)) + \|r_1 p_2\|^2 + \mathrm{cost}(\pi^b(T_2)) + \ldots + \|r_{k-1} p_k\|^2 + \mathrm{cost}(\pi^b(T_k)) \\
&\leq (\|rr_1\| + \|r_1 p_1\|)^2 + \mathrm{cost}(\pi^b(T_1)) \\
&\quad + (\|r_1 r\| + \|rr_2\| + \|r_2 p_2\|)^2 + \mathrm{cost}(\pi^b(T_2)) \\
&\quad \vdots \\
&\quad + (\|r_{k-1} r\| + \|rr_k\| + \|r_k p_k\|)^2 + \mathrm{cost}(\pi^b(T_k)) \\
&\leq 2\|rr_1\|^2 + 2\|r_1 p_1\|^2 + \mathrm{cost}(\pi^b(T_1)) \\
&\quad + 3\|r_1 r\|^2 + 3\|rr_2\|^2 + 3\|r_2 p_2\|^2 + \mathrm{cost}(\pi^b(T_2)) \\
&\quad \vdots \\
&\quad + 3\|r_{k-1} r\|^2 + 3\|rr_k\|^2 + 3\|r_k p_k\|^2 + \mathrm{cost}(\pi^b(T_k)) \\
&\leq 6 \sum_{i=1}^{k} \|rr_i\|^2 + 3 \sum_{i=1}^{k} \|r_i p_i\|^2 + \sum_{i=1}^{k} \mathrm{cost}(\pi^b(T_i)) - 3\|rr_k\|^2 \\
&\leq 6 \sum_{i=1}^{k} \|rr_i\|^2 + 6 \sum_{i=1}^{k} \mathrm{MST}(T_i) - 3\|rr_k\|^2 \\
&= 6\,\mathrm{MST}(T) - 3\|rr_k\|^2.
\end{aligned}
$$

In the above calculation we used the fact that $(\sum_{i=1}^{n} a_i)^\alpha \leq n^{\alpha-1} \cdot \sum_{i=1}^{n} a_i^\alpha$, for $a_i \geq 0$ and $\alpha \geq 1$. The path $\pi^b(T)$ is constructed analogously.

In fact, the very same construction and reasoning can be generalized to the following corollary.

**Corollary 7** *There exists a $2 \cdot 3^{\alpha-1}$-approximation algorithm for the TSP if the edge weights are Euclidean edge weights to the power $\alpha$.*